\documentclass[epsfig,12pt]{article}
\usepackage{epsfig}
\usepackage{graphicx}
\usepackage{amsmath,amsfonts,mathtools,bbold,amssymb,color,relsize}
\usepackage{tikz-cd}
\usepackage[margin=1.25in]{geometry}

\newcommand{\beq}{\begin{equation}}   
\newcommand{\eeq}{\end{equation}}
\newcommand{\beqn}{\begin{equation}\left.\begin{aligned}}   
\newcommand{\eeqn}{\end{aligned}\right.\end{equation}}

\newcommand{\Int}{\int\limits}
\newcommand{\Z}{\mathbb{Z}}
\newcommand{\R}{\mathbb{R}}

\newcommand{\phiq}{\varphi_{\text{qu}}}
\newcommand{\xq}{x_{\text{qu}}}
\newcommand{\theq}{\theta_{\text{q}}}
\newcommand{\theo}{\theta_{\text{0}}}

\newcommand{\gsim}{\lower.7ex\hbox{$
\;\stackrel{\textstyle>}{\sim}\;$}}
\newcommand{\lsim}{\lower.7ex\hbox{$
\;\stackrel{\textstyle<}{\sim}\;$}}

\begin{document}

\begin{titlepage}

\begin{flushright}
FTPI-MINN-23-14\\
UMN-TH-4222/23
\end{flushright}

\vspace{7mm}

\begin{center}
{  \bf{\large   $U(1)$ Defects on Domain Lines}}
\end{center}

%\vspace{1mm}

\begin{center}

 {\large 
 Evgeniy Kurianovych$^a$ and Mikhail Shifman$^{a,b}$}
\end {center}

\begin{center}

$^{a}${\em School of Physics and Astronomy, 
 University of Minnesota,
Minneapolis, MN 55455, USA}

 \vspace{2mm}
 
$^b${\em William I. Fine Theoretical Physics Institute, University of Minnesota,
Minneapolis, MN 55455, USA}

\end {center}

%%%%%%%%%%%%%%%%%%%

\vspace{10mm}

%%%%%%%%%%%%%%%%%%%
\begin{center}
{\large\bf Abstract}
\end{center}

\hspace{0.3cm}
Based on recent experimental results, we give field-theoretic description of $U(1)$ defects localized on the domain lines on thin films. We describe topology of our model and solve this model in the adiabatic approximation. It turns out that such a model naturally provides periodic structure observed in experiment. The effective theory turns out to be the sine-Gordon model, but unlike the previous theoretical considerations we argue that in this case it is favorable for sine-Gordon kinks to merge into one defect with a uniform winding. We consider a system of adjacent domain lines and anti-lines and explain the experimental fact that the appearance of defects on a domain line prevents defect creation on the adjacent anti-lines. We also quantize the model and investigate possible effects of finite transverse dimension of the film.
	
\vspace{2cm}

\end{titlepage}

\newpage

\section{Introduction}

The study of Skyrmions has a long history. Its beginning was initiated in the 1960s by Skyrme \cite{Skyrme}. The subsequent development due to Witten et al. followed \cite{Witten}. The Skyrme model provides quite a good description of some properties of baryons. Similar models also exist in condensed matter systems (see e.g. \cite{SchSh}), but the condensed matter applications usually involve baby Skyrmions\footnote{Technically baby Skyrmions are the Polyakov-Belavin instantons \cite{Polyakov} in 2d $CP(1)$ model in the continuous limit.}. The difference is that while in nuclear physics Skyrmions exist in 3 spatial dimensions and are based on topologically non-trivial maps from $\R^3$ (with points at infinity identified) to $S^3$, baby  Skyrmions live on a spatial plane and are based on topologically non-trivial maps from $\R^2$ to $S^2$.

In the recent years topological solitons in condensed matter attracted a considerable experimental effort \cite{Tokunaga,Nayak,DasTang,NagaseSmectic,GaoJe,TaeHoonKim,Villalba,ChengJieWang}. It is driven by the perspective of using topological solitons in spintronic and memory devices. Sometimes those solitons are denoted by a common word Skyrmions, although there is a vast number of different systems that show rich internal structure. In particular, it is interesting to investigate how different topological defects can be combined together.

Different approaches can be applied to combine a Skyrmion and a domain wall. Our previous method \cite{Sh1,KurSh,BKK} used two coupled fields. One of them condenses in the vacuum and gives rise to a domain wall. For another one it is energetically favorable to condense only on the wall, and its condensation provides the internal structure necessary to make a Skyrmion. Such an approach is simple and convenient, but the underlying structure of real condensed matter systems is quite different.

Also there exists an approach of \cite{GudnasonNitta}. It uses putting a 3d Skyrmion into a domain wall. In this way it is possible to have no Skyrme term that is usually needed to stabilize Skyrmion size, since such a stabilization is performed by the domain wall itself. But such an approach might be challenging to realize in the condensed matter applications.

Here we study the experimental results of \cite{Nagase} which combine defects (which the authors call bimerons) with domain walls on thin films. We investigate the topology of the system and describe this system in the continuous limit. We use the standard way of describing magnetic systems in field theory, which is the $O(3)$ sigma model. We also need other ingredients: magnetic anisotropy, which in necessary to create a domain line, and the Dzyaloshinsky-Moria (DM) interaction which winds the magnetic moments.

In section \ref{TopologySection} we qualitatively describe the model and its topology. It turns out that the domain line appears between two vacua in the two poles of the target space sphere. There is a freedom as to along which meridian these two poles are connected. This gives rise to a $U(1)$ modulus localized on a domain line. Winding the real space domain line around this target space $U(1)$ creates the chains of defects that are observed in experiment. In section \ref{TwistedSection} we review the sigma model with a "twisted mass", which is relevant to the system under consideration and convenient due to being solvable analytically. In section \ref{AdiabaticSection} we solve the model in the adiabatic approximation, assuming that the DM interaction is a small perturbation in the twisted mass model. The effective model turns out to be the sine-Gordon theory, but unlike the previous theoretical considerations we argue that in this case it is favorable for the sine-Gordon kinks to merge into one defect with a uniform winding. In section \ref{ModuliSection} we consider small quantum fluctuations in this system. In section \ref{CrystalSection} we consider several parallel domain lines and anti-lines and explain the appearance of alternating lines with and without defects. In section \ref{StabilitySection} we discuss the stability of the system, related to finite width of the film.

A similar model was considered in \cite{RossNitta} (see also \cite{Nitta4} and \cite{Nitta5}). Unlike \cite{RossNitta} we argue that having  winding with a constant rate might be energetically more favorable than the chain of separate sine-Gordon defects\footnote{Intermediate results, in particular on the uniform winding, were discussed with one of the authors of \cite{RossNitta} (Muneto Nitta) in June 2023.}. We also consider crystals of defects and quantization of the model,  which is absent in \cite{RossNitta}.

\section{Qualitative features of the model and its topology}\label{TopologySection}

\begin{figure}
      \epsfxsize=450px
   \centerline{\epsffile{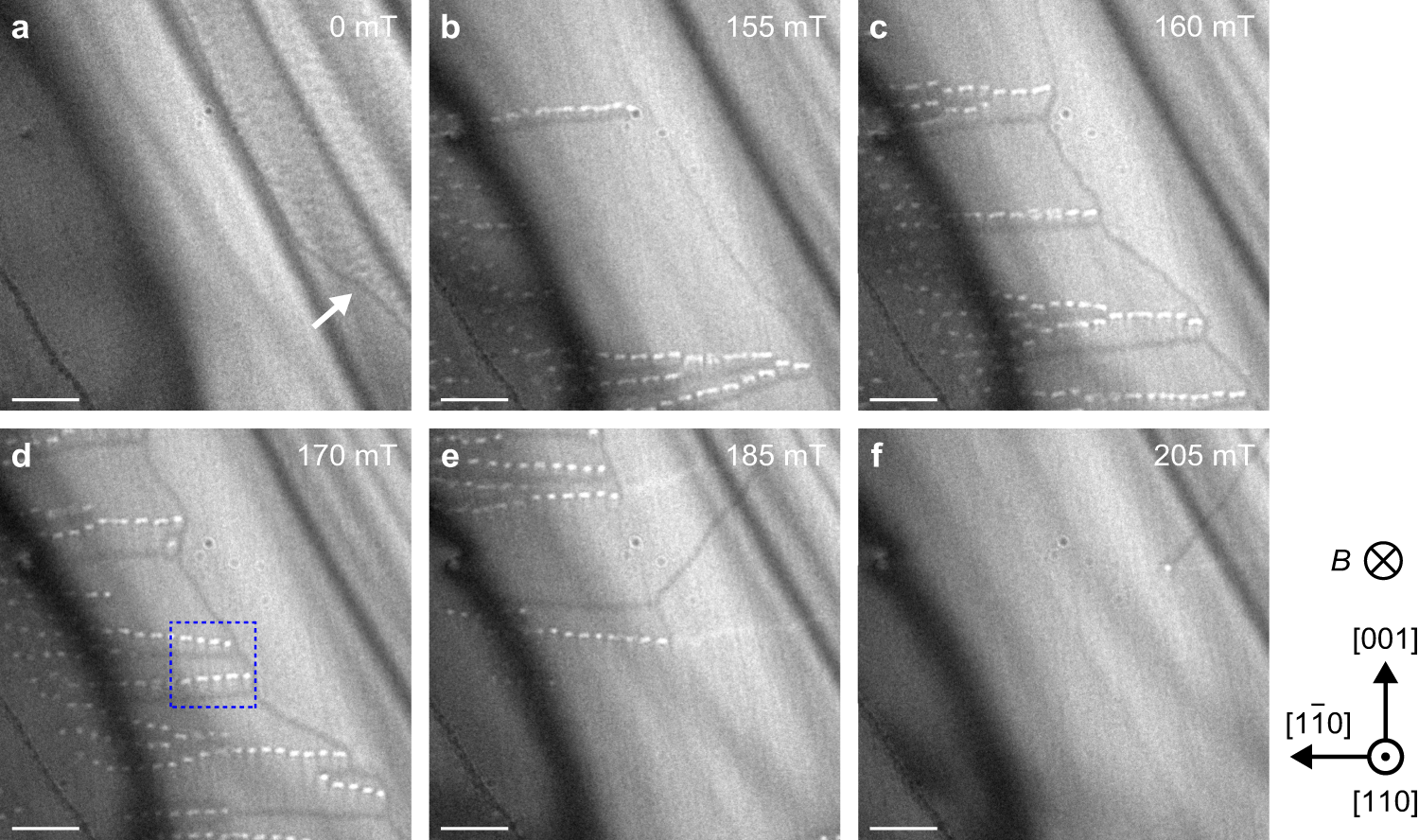}}
   \caption{Experimental results from \cite{Nagase}. The bright dashes are defects localized on domain lines in thin films. The dark lines are domain lines with no defects on them. These two types of lines are alternating, and lines of the same type never appear next to each other. The images were obtained using the Lorentz transmission electron microscopy (LTEM) which shows magnetization. The scale bars are 1 $\mu$m.}
   \label{n1}
\end{figure}

The experimental results from \cite{Nagase} are shown in Fig. \ref{n1}. We see that the defects are localized on domain walls and appear in chains. There are alternating walls with and without defects on them. To find the best theoretical description, let us consider the qualitative properties of this experimental setup. The experiment was performed in thin films whose width was smaller than the topological defect size. So it was effectively a 2d system\footnote{Strictly speaking, it is a system in 2+1 dimensions, but now we consider a static limit. Possible time dependence will be investigated in section \ref{ModuliSection}.}, and defects were localized on the domain lines. A natural description of magnetic moments is provided by the $O(3)$ sigma model, where the target space vectors have three real coordinates. Those vectors have fixed lengths, i.e. are localized on a sphere $S^2$.

To create a domain line we need magnetic anisotropy to break the $O(3)$ symmetry. This is supported by a potential which has two minima on the two poles of the target space sphere $S^2$. In vacuum the magnetic moment picks one of these two directions. If at $x=+\infty$ the system is in one vacuum while at $x=-\infty$ it is in the other vacuum, the transition region will provide us with a domain line.

Stability of the topological defect under consideration requires another ingredient. Its stability is ensured on topological grounds, since for a topologically non-trivial system there is no continuous transformation that would transform the system in a state with no defect. But such a system is just quasistable, since the state without a defect has a lower energy. Therefore in a real condensed matter system such a defect will inevitably shrink. So topological stability is not sufficient, and to make a defect stable with respect to its size we need a system where the state with a defect has the lowest possible energy. This is achieved via the  Dzyaloshinskii-Moriya (DM) interaction with the energy:
\beq\label{HDM}
H_{DM} = \sum_{i,j} {\bf C}_{i,j} \cdot ({\bf S_i}\times{\bf S_j}).
\eeq
Here the indices $i$ and $j$ label atomic sites and ${\bf C}_{i,j}$ are some constants responsible for the strength and spatial structure on the interaction. Such an interaction makes it favourable for spins to be orthogonal to each other and in this way "winds" them.
\begin{figure}
      \epsfxsize=350px
   \centerline{\epsffile{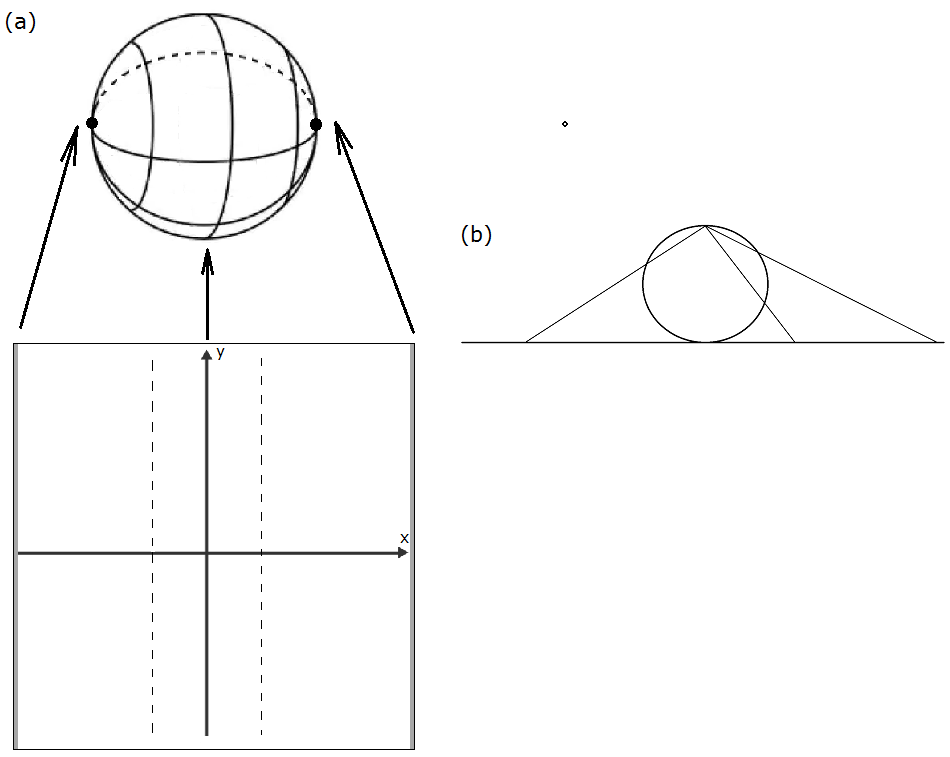}}
   \caption{(a) Mapping of the real plane to the target space $S^2$. The poles of the target space $S^2$ are denoted by solid black circles $\bullet$. These poles correspond to the vacua of the system. The domain line (dashed) is centered at the $y$ axis. The middle of the domain line is mapped onto the target space equator while all the points at $x=\pm\infty$ are mapped onto the poles. The lines parallel to the $x$ axis are mapped onto the target space meridians. (b) Mapping $g$ of the $y$ axis to the equator of the target space $S^2$.}
   \label{Spheremapfig}
\end{figure}

Let us summarize. We have an $O(3)$ sigma model in 2+1 dimensions with a potential that breaks this $O(3)$ symmetry down to $\Z_2$ symmetry of the vacuum manifold. Picking one of the vacua breaks this $\Z_2$. A region that interpolates between the two vacua is a domain line. Since the two vacua correspond to the two poles of the target space sphere $S^2$, the intermediate region between them corresponds to the equator of that sphere as depicted in Fig. \ref{Spheremapfig} (a). A freedom to pick an arbitrary point on that equator leads to the emergence of a $U(1)$ modulus localized on the domain line. We are investigating the defects that appear from this modulus.

If we do not have a DM term then the domain line with the lowest energy would correspond to an arbitrary target space trajectory along a meridian between the $S^2$ poles. Now assume that the DM interaction is non-zero, but small. Then in the bulk it is suppressed and magnetic moments are just parallel to each other. But in the domain line region we have non-zero field derivatives, so the DM term makes it energetically favourable to wind the target space position around the equator while we go along the domain line. We will investigate the solution this model in the next sections. Now let us discuss its topology.

Usually "baby" Skyrmions in 2d are based on a map from the space with all points at the infinity identified, so we can treat it topologically as a sphere. But having a domain line makes such an identification impossible.

Assume that the domain line goes along the $y$ axis. Then what we need is a topologically non-trivial map $f: \R^2\rightarrow S^2$ such that for an arbitrary $y$ we have $f(-\infty,y)$ going to the north pole while $f(+\infty,y)$ goes to the south pole.

This can be done in the following way. First we map the $y$ axis to the equator as shown in Fig. \ref{Spheremapfig} (b). The points $y=\pm\infty$ are mapped to the same point on the circle, so this map wraps the $y$ axis around the equator. Then we map the lines parallel to the $x$ axis to the meridians of $S^2$, starting from the relevant point on the equator. Let us use, for instance, the arctan function. If we measure the target space latitude $\psi$ from the equator then
\beq
\psi(x,y) = -\arctan(x).
\eeq
So the overall map in spherical coordinates (with 0 latitude on the equator) is $f: (x,y)\rightarrow(g(y),-\arctan(x))$, where the function $g$ is defined by Fig. \ref{Spheremapfig} (b). Such an $f$ is obviously topologically non-trivial. Of course, it is not an exact solution, but it shows the desired topological configuration.

It is interesting that in this case we do not have separate defects, but they create a continuous chain. This naturally follows from the chosen topology, since (unlike the usual baby Skyrmions where all the points at infinity should be in the same vacuum, therefore enforcing the finite size) now we can wind the $y$ axis around the equator of $S^2$ arbitrarily many times. This is clearly supported by the experimental results (see Fig. \ref{n1}).

It should be noted that in experiment we see a continuous chain of separate defects and not one continuous defect. But such a picture might actually happen due to the detection technique. Since it observes magnetization along specific directions, some directions might appear invisible. So when magnetization winds around the target space equator, some parts of that equator will not be detected. We will further discuss this question in the next sections.

Depending on the boundary conditions the topological charge can also be non-integer. The general expression for the topological charge is given just by the integral of the target space polar angle $\theta$:
\beq
\int d\theta = \int \frac{\chi_2 d\chi_1-\chi_1 d\chi_2}{\chi_1^2+\chi_2^2} .
\eeq
Here $\chi$ is the magnetic moment ${\bf S}$ in the continuous limit (see below, section \ref{TwistedSection}). Such an integral is topologically invariant and can be calculated along any line that does not touch the poles in the target space.

\section{Twisted mass without DM}\label{TwistedSection}

In field theoretical language the magnetic moment ${\bf S_i}$ at the site $i$ corresponds to a continuous three-component real target space function $\lbrace\chi_i\rbrace, i=\lbrace1,2,3\rbrace$. It has a constant magnitude $a$, i.e it is subject to constraint
\beq
\chi_i\chi_i = a^2.
\eeq
We do not distinguish the upper and lower target space indices. It is convenient to introduce normalized functions
\beq
S_i\equiv\chi_i/a.
\eeq
For brevity the three component target space vector will be written sometimes as $\overrightarrow{S}$ or $\overrightarrow{\chi}$. Then the action of a sigma model in $D$ dimensions with a coupling constant $g$ is
\beq\label{O3act}
S = \frac{1}{2g^2} \int d^Dx (\partial_\mu S_i)(\partial^\mu S_i).
\eeq
It is convenient to write this model in complex coordinates. Vector $\overrightarrow{S}$ is localized on a two-dimensional sphere $S^2$ which can be mapped onto a plane via a stereographic projection. If $\phi_1$ and $\phi_2$ are coordinates on the complex plane then
\beq
S_1 = \frac{2\phi_1}{1+\phi_1^2+\phi_2^2}, \quad
S_2 = \frac{2\phi_2}{1+\phi_1^2+\phi_2^2}, \quad
S_3 = \frac{1-\phi_1^2-\phi_2^2}{1+\phi_1^2+\phi_2^2}.
\eeq
We can treat real fields $\phi_1$ and $\phi_2$ as components on one complex field $\phi$:
\beq
\phi_1(x) \equiv \text{Re}\hspace{0.7mm}\phi(x), \quad
\phi_2(x) \equiv \text{Im}\hspace{0.7mm}\phi(x)
\eeq
The inverse transformation from $\overrightarrow{S}$ to $\phi$ is given by
\beq
\phi = \frac{S_1+iS_2}{1+S_3}.
\eeq
So in two dimensions we get the action
\beq\label{TWact}
S_{TW} = \int d^2x \hspace{1mm} G \hspace{0.7mm}(\partial_\mu\bar{\phi}\partial^\mu\phi - |m|^2\bar{\phi}\phi),
\eeq
where
\beq
G = \frac{2}{g^2(1+\bar{\phi}\phi)^2}.
\eeq
The first term in \eqref{TWact} is just the action \eqref{O3act} written in the complex coordinates. The second term that we added here is called a twisted mass term. In terms of \eqref{O3act} it is
\beq\label{twistedm}
m^2(1-S_3^2).
\eeq
It is a potential that gives us what we need from the magnetic anisothropy: it has minima at the poles of the target space $S^2$ sphere. These minima are the two vacua corresponding to $\phi=0$ and $\phi=\infty$.

Let us find the solution for a kink that connects these two vacua. This can be done from a BPS equation. Assuming no time dependence we rewrite the action \eqref{TWact} as
\beq
- \left[ \int dx \hspace{0.8mm} G \hspace{0.5mm} (\partial_x\bar{\phi}-|m|\bar{\phi})(\partial_x\phi-|m|\phi)+|m|\int dx \partial_x h \right],
\eeq
\beq
\partial_x h \equiv G (\phi\hspace{0.5mm}\partial_x\bar{\phi}+\bar{\phi}\hspace{0.5mm}\partial_x\phi).
\eeq
Then the BPS equation reads
\beq
\partial_x\phi = |m|\phi,
\eeq
and its solution is \cite{Shbook}
\beq\label{TWsol}
\phi(x) = e^{|m|(x-x_0)-i\beta}.
\eeq

This solution was obtained for a time-independent configuration of a one-dimensional kink in 1+1 dimensional theory. But the same solution can be used in 2 spatial dimensions. Instead of $x$ dependent configuration that does not depend on time it will be an $x$ dependent configuration that does not depend on $y$, i.e. a domain line.

A free parameter $\beta$ reflects the fact that the trajectory between the $S^2$ poles can go along any meridian. The phase $\beta$ is defined on a $U(1)$ circle. This is exactly what we need to build the $U(1)$ defects. Now it is an arbitrary parameter, but in equilibrium it should have the same value everywhere. If we add the DM interaction to the model, it will actually wind around this $U(1)$ in the target space while we go along the domain line in the real space.

\section{Adiabatic approximation}\label{AdiabaticSection}

Now we are ready to consider the full model. Its Lagrangian is
\beq
{\cal L} = \frac{1}{2} \partial_\mu\chi_i \partial^\mu\chi_i - 
\lambda(\chi_i\chi_i-a^2)^2 - V(\chi_i) - V_{\text{DM}}.
\eeq
Here $V_{\text{DM}}$ is the DM term. Its specific form is discussed below. We assume $\lambda$ to be large, so the term proportional to $\lambda$ just imposes the constraint on the value of $\chi^i\chi^i$. The potential term $V$ reflects the anisotropy and makes the $\chi^3$ direction preferable. It can be either quartic
\beq\label{quarticV}
V_4 = \varkappa(a^2 - \chi_3^2)^2,
\eeq
or quadratic twisted mass term:
\beq
V_2 = \eta(a^2-\chi_3^2)
\eeq
The potential $V_2$ is just the twisted mass potential \eqref{twistedm}. It is somewhat more convenient than \eqref{quarticV}, since we can use the analytic solution \eqref{TWsol}. The quartic potential $V_4$ is more relevant to experiment. But the qualitative behaviour of the system is the same in both cases, since it requires only a potential with minima at the poles and the DM term. The way of adiabatically solving the model is also applicable for the $V_4$ potential, but instead of the solution \eqref{TWsol} we would need to use a numerically calculated profile of the theory without DM interaction.

We assume the film to be thin, so the fields do not change along the $z$ direction. Therefore it can be integrated out to give
\beq
S = L \int {\cal L}(x,y,t) dx dy dt.
\eeq
Here $L$ is the film thickness. Such a theory is effectively (2+1)-dimensional, but has parameters inherited from the original (3+1) dimensional theory.

For the DM term there can be different possibilities. The expression \eqref{HDM} for discrete spins describes the interaction of spins near each other. Each spin interacts with several different neighbors. So to write an expression for the Lagrangian density in continuous limit we need to perform averaging over all these neighbors. The result will depend on the specific form of the constants ${\bf C}_{i,j}$ which depend on the material properties. The simplest expression for the DM term in the continuous limit is
\beq\label{DMspher}
C^{k \mu} \varepsilon_{ijk} \chi_i \partial_\mu \chi_j.
\eeq
Straightforward identification of the spatial and target space indices can be realized by $C^{k\mu}=c\delta^{k\mu}$. But for thin films the more relevant form might be \cite{Cortes}:
\beq
c\left[\overrightarrow{\chi}\cdot\nabla\chi_3 - \chi_3(\nabla\overrightarrow{\chi})\right] =
c(\chi_i\partial_i\chi_3 - \chi_3\partial_i\chi_i),
\eeq
where $c$ is some constant describing the strength of the DM interaction.

We solve the model in the adiabatic approximation. We assume that the DM term is small, and so the difference between the solution \eqref{TWsol} and the perturbed solution is also small.

It is convenient to pass to the spherical coordinates $(\varphi,\theta)$ in the target space:
\beq
\chi_1 = a\sin\theta\cos\varphi, \qquad \chi_2 = a\sin\theta\sin\varphi, \qquad
\chi_3 = a\cos\theta.
\eeq
As before, we assume that interpolation between the vacua at $x=\pm\infty$ is associated with the domain line along the $y$ axis. In these coordinates the solution \eqref{TWsol} means that the angle $\varphi$ remains constant everywhere and we know the angle $\theta$ as a function of $x$. Let's call this function $\theo(x)$.

If we turn on a small DM term, the angle $\varphi$ will vary slowly as a function of $x$ and $y$. The angle $\theta$ will receive a small correction $\theq$. But if the DM term is spherically symmetric, the family of the domain line solutions should be invariant under the $U(1)$ action. This means that if we rotate some solution by an arbitrary angle around the target space axis between the two poles, the result should also be a possible solution. This is possible only if the perturbed $\theta$ does not depend on $y$ and $\varphi$ can be split into separate parts depending on $x$ and $y$:
\beqn
&\varphi(x,y) = \varphi_1(x)+\varphi_2(y), \\
&\theta(x,y) = \theo(x) + \theq(x).
\eeqn

\begin{figure}
      \epsfxsize=250px
   \centerline{\epsffile{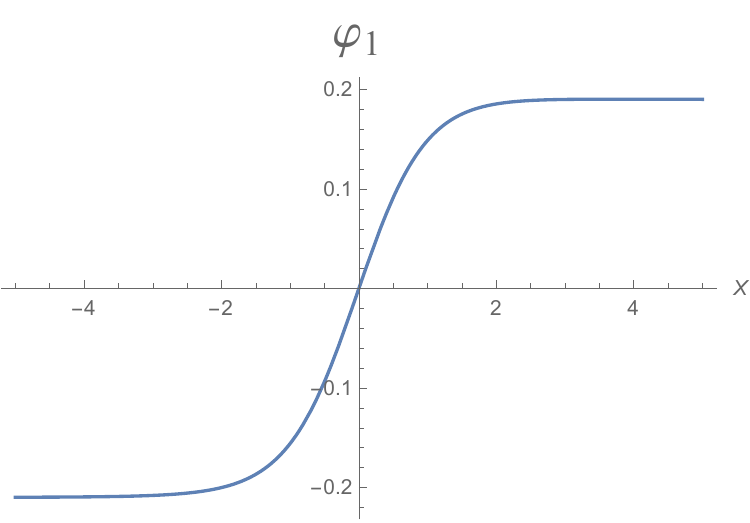}}
   \caption{Numerical solution for $\varphi_1(x)$. Here $c/m=0.01$ and the units for $x$ are $1/m$.}
   \label{profilepic}
\end{figure}

First we determine $\varphi_1(x)$. To this end we expand the Lagrangian up to the second order in $\theq$ and $\partial\varphi$. Since $\theo$ minimizes the action without the DM term, so there will be no linear terms in that part of the action. Since $\varphi_2(y)$ varies slowly and $\theta_q\ll\theta_0$, we can neglect $\theta_q$ and $\varphi_2$ here. In this way we get a differential equation for $\varphi_1$ in one variable $x$:
\beq
\varphi_1'' + 2\theo'\varphi_1'\cot\theo + c\theo' = 0.
\eeq
We solve it numerically. Without loss of generality we can set $\varphi_1(0)=0$. Since $\varphi_1'(0)$ is unknown, we used shooting method, varying that derivative until we found a solution that is finite at infinity and either even or odd with respect to $x$. Non-zero derivative at $x=0$ implies an odd solution. This is reasonable, since otherwise we would have a maximum or minimum at $x=0$. But a non-constant $\varphi$ is created by $\theta_0'$ which has no extremal value at $x=0$. The numerical solution for $\varphi_1(x)$ is depicted in Fig. \ref{profilepic}.

In the adiabatic approximation we assume that the winding along the $y$ direction is slow. Quantitatively this means that the characteristic length on which $\varphi$ changes significantly in the $y$ direction is much larger than the domain line width in the $x$ direction. So if we look for $\varphi_2$ we can integrate out the $x$ direction to get an effective action for $\varphi_2(y)$. Since $\theta_q\ll\theta_0$ we can neglect $\theta_q$ here.

When we perform the $x$ integration we need to take into account the symmetry properties. We know that the polar angle with respect to the equatorial plane
\beq
\psi=\dfrac{\pi}{2}-\theta_0
\eeq
is an odd function: $\psi(-x)=-\psi(x)$ Therefore $\theta_0'(x)=\theta_0'(-x)$. Also we found that $\varphi_1(-x)=-\varphi_1(x)$. Using these properties, we get the following 1d effective Lagrangian for the $y$ coordinate:
\beq\label{1dEffLagr}
{\cal L}_{1d}(y) = -\rho_0(\varphi_2')^2 - c\rho_1\cos\varphi_2 - 
c\rho_2 \varphi_2' \sin\varphi_2,
\eeq
where
\beqn\label{1dparameters}
\rho_0 &= \Int_{-\infty}^{\infty} \sin^2\theta_0(x) dx. \\
\rho_1 &= \Int_{-\infty}^{\infty} \left[\theta_0'(x) \cos\varphi_1(x)+
\frac{1}{2}\varphi_1'(x)\sin\varphi_1(x)\sin 2\theta_0(x)\right] dx, \\
\rho_2 &= \frac{1}{2}\Int_{-\infty}^{\infty} \sin 2\theta_0(x) \sin\varphi_1(x) dx.
\eeqn
The last term in \eqref{1dEffLagr} disappears in the equation of motion. So we get just a sine-Gordon model, similarly to \cite{RossNitta}. The equation of motion is
\beq\label{SGequation}
\varphi_2'' - c\frac{\rho_1}{2\rho_0} \cos\varphi_2 = 0.
\eeq
Its solution is well known:
\beq
\varphi_2(y) = 4\arctan\exp\left[\sqrt{\frac{c\rho_1}{2\rho_0}}(y-y_0)\right],
\eeq
where $y_0$ is a free parameter describing the center of the defect.

Such a solution implies a localized defect. This is the result used in \cite{RossNitta}. Several defects of the same type then can appear next to each other. It is somewhat different from what we described in Chapter 2. There, we had the same topology, but in a continuous setup. Now the defects are separate. The actual result would depend on which configuration has the lowest energy.
\begin{figure}
      \epsfxsize=250px
   \centerline{\epsffile{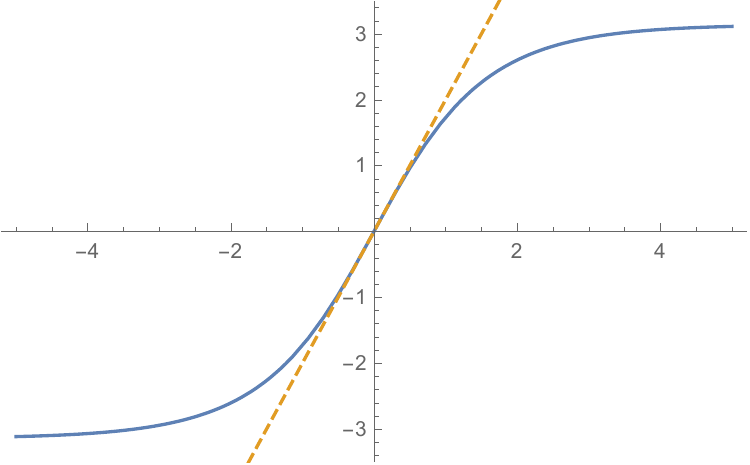}}
   \caption{Plot of the function $(4\arctan e^y-\pi)$ (solid) and its derivative at zero (dashed). We see that the profile of the solution is close to linear for a significant part of its size.}
   \label{Arctan}
\end{figure}

\begin{figure}
      \epsfxsize=380px
   \centerline{\epsffile{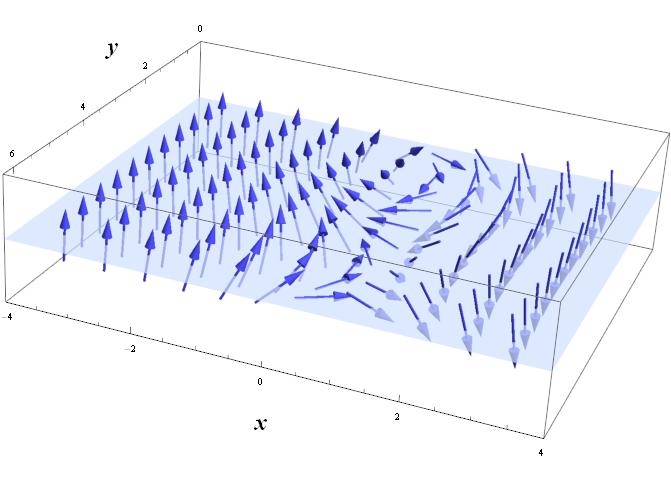}}
   \caption{Profile of a $U(1)$ defect. The domain line runs along the $y$ axis, and the magnetic moment rotates around the target space equator along it. The image is not up to scale, since in the adiabatic approximation the winding is slow. So the distance along the $y$ axis corresponding to a $2\pi$ rotation should be much larger than the one depicted here. The non-uniform winding described above is shown in Fig. \ref{profilepic}.}
   \label{Arrows1}
\end{figure}

We argue that the continuous configuration might actually be energetically favorable. Indeed, creation of a defect itself is energetically favorable. But most of the defect energy is concentrated near its center. So the most energetically favorable configuration would be the one which is closest to the defect profile near its center, provided that it can actually exist. If we look at the profile function on Fig. \ref{Arctan}, we see that it is linear in a large region near the center. But this is exactly the uniform winding that we have described above, given by the equation
\beq
\varphi_2(y) = \alpha (y-y_0), \qquad \alpha = \sqrt{\frac{2c\rho_1}{\rho_0}}.
\eeq
The structure of a defect with a uniform winding in $y$ direction and with $\theta_0$ defined by  \eqref{TWsol} is depicted in Fig. \ref{Arrows1}.

If we are dealing with the adiabatic approximation we assume that $\varphi_2(y)$ changes slowly. But then in the linear region it will not feel the end of the defect and can be continued for an arbitrary distance. Such a solution would restore the translational symmetry broken by a solitary kink. It should also be noted that the Lagrangian of the system has the translational symmetry unbroken. This is unlike the generic sine-Gordon model, where the potential with the cosine function explicitly breaks this symmetry.

The experimental data are not conclusive in that respect. In Fig. \ref{n1} we see that the distance between domain lines is comparable with the defect size, so the adiabatic approximation is not applicable here. And, as was noted above, the gaps between defects on the image do not necessarily mean that there is an actual gap. The other argument supporting that we do not have just an ordinary sine-Gordon model is that the usual sine-Gordon kinks repel each other \cite{Manton}. But the defects under consideration appear in experiment as tight chains with a small distance between the defects. More relevant here may be the results on  inter-skyrmion attraction in 2d chiral magnets with in-plane anisotropy \cite{Kameda}. Another possibility in our case would be a partial fusion of defects. In this case the adjacent defects will overlap significantly, but will not reach the completely uniform winding.

If the DM term is given by \eqref{DMspher}, the 1d effective action is
\beq\label{1dEffLagrSpher}
{\cal L}_{S1d}(y) = -\rho_0(\varphi_2')^2 + c\rho_1\sin\varphi_2 + 
c\rho_2 \varphi_2' \cos\varphi_2.
\eeq
It is essentially the same as \eqref{1dEffLagr}, since sine and cosine can be turned to each other just by changing the initial value of $\varphi_2$. The coefficients $\rho_{0,2}$ are given by the same expressions as in \eqref{1dparameters} and
\beq
\rho_1 = \Int_{-\infty}^{\infty} \left[\theta_0'(x) \cos\varphi_1(x)-
\frac{1}{2}\varphi_1'(x)\sin\varphi_1(x)\sin 2\theta_0(x)\right] dx.
\eeq
We denoted the DM coupling constant in both cases by the same letter $c$.

If the DM term is not spherically symmetric the coefficient $c$ will be different for different regions of the target space sphere. In the adiabatic approximation this would lead to a slowly changing function $\alpha(y)$.

\section{Moduli and quantization}\label{ModuliSection}

If we look at the small deviations from the profile above we will get a (1+1)-dimensional effective action. We have two moduli. The first is the polar angle $\varphi_2$ which now has a quantum correction to the solution given above:
\beq
\varphi_2(t,y) = \alpha y +\phiq(t,y).
\eeq
The second is the position of the line center which now depends on $y$ and $t$: $\xq(t,y)$. As before, the unperturbed solution is centered at $x=0$. Since the deviations are small, the profile functions $\theo$ and $\varphi_1$ remain unchanged, but can shift as a whole depending on $x-\xq(t,y)$:
\beqn
&\theta(t,x,y) = \theo(x-\xq(t,y)), \\
&\varphi(t,x,y) = \alpha y + \varphi_1(x-\xq(t,y)) +\phiq(t,y).
\eeqn

Now we plug this solution into the original Lagrangian. Some of the terms can be neglected in the effective action. These are (a) the terms already minimized by the solution. (b) Linear terms that vanish, since they are first order corrections to the extremal action. (c) Some potential terms, since they are not affected by the translational modulus $\xq(t,y)$, which matters only when we have derivatives. (d) The DM term cancels the $\alpha y$ in the $\varphi$, since this $\alpha y$ is the solution in presence of the DM interaction, and we are looking for the corrections to that solution. So we are left with the following effective action:
\beq\label{Seff}
S_{\text{eff}} = a^2 L \Int dt dy \left[ I_1 (\partial\xq)^2 - 
I_2(\partial_p\xq \partial^p\phiq + \partial_p\phiq\partial_p\xq) +
I_3(\partial\phiq)^2 \right], \quad p=t,y.
\eeq
We kept the cross terms separate in view of quantization. Here
\beqn
I_1 &= \Int \left[(\varphi_1')^2 \sin^2 \theo +(\theo')^2 \right]dx, \\
I_2 &= \Int \varphi_1' \sin^2 \theo dx, \\
I_3 &= \Int \sin^2 \theo dx. 
\eeqn

From the classical effective action we can proceed to quantization. We promote $\xq$ and $\phiq$ in\eqref{Seff} to operators and diagonalize it by introducing
\beq
\hat{\xi}_{\text{qu}} = \sqrt{I_1}\hat{x}_{\text{qu}} - \frac{I_2}{\sqrt{I_1}}\hat{\varphi}_{\text{qu}}.
\eeq
Then we have quanta of the two massless fields $\hat{\xi}_{\text{qu}}$ and $\hat{\varphi}_{\text{qu}}$. If we add higher order terms in the Taylor expansion the interactions will appear.

Since the underlying structure is periodic with the length
\beq
\frac{2\pi}{\alpha},
\eeq
we can expect some kind of resonance for such a wavelength.

\section{Crystals of $U(1)$ defects}\label{CrystalSection}

\begin{figure}
      \epsfxsize=270px
   \centerline{\epsffile{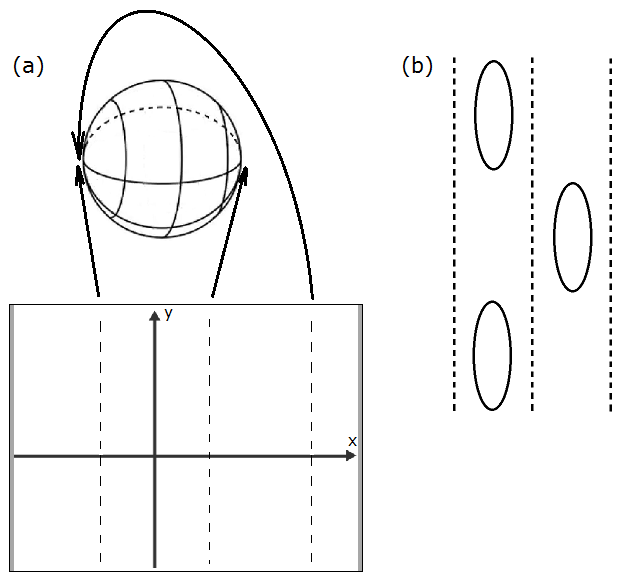}}
   \caption{(a) A domain line and adjacent anti-line. The left dashed line corresponds to the vacuum at the north pole and the middle line corresponds to the vacuum at the south pole of the target space sphere. The region between them is the domain line. The right dashed line is mapped to the north pole and the region between the middle and right dashed lines is the domain anti-line. (b) A theoretically possible configuration where a defect on the anti-line fits the space between two defects on the domain line. An ellipse corresponds to $2\pi$ winding around the target space equator. Absence of configurations of that type in experimental results proves that there are no significant gaps between defects of domain lines, so the system of defects is always a continuous one.}
   \label{Spheremapfig2}
\end{figure} 

Let us consider several domain lines adjacent to each other, as they appear in the experiment \cite{Nagase}. In the state of perfect equilibrium the different lines would be just independent of each other. If we go in the direction perpendicular to the lines, profile of one line will reach vacuum and the next line will appear after that. Different domain lines can overlap with each other. Such an overlap would bring extra energy. If the system with no overlap is stable, it is protected by some energy barrier. Therefore if the extra energy provided by the overlap is smaller than that barrier, the system would remain stable. Such an overlap would lead to repulsion between lines, but if the line motion is very slow or prevented by the film boundary, such a system can be considered as quasistable.

Let us consider two adjacent domain lines parallel to the $y$ axis (Fig. \ref{Spheremapfig2}). If we go along the $x$ axis, the mapping to the target space goes from the north pole on the left to the south pole on the right. Then the anti-line starts, when we go from the south pole to the north pole. How would the configurations with several domain lines affect the defects on these lines?

As was described above, winding around the meridian happens due to the $\theo$ derivative in the $x$ direction. If instead of a domain line we have an anti-line, then the $\theo$ derivative would have an opposite sign. Then the parameter $\rho_1$ in \eqref{1dparameters} will also have an opposite sign, since the same holds for $\sin2\theo$. So the second term of the sine-Gordon equation \eqref{SGequation} will have an opposite sign as well. This is equivalent to a transformation $\varphi\rightarrow\varphi+\pi$ in \eqref{SGequation}. Therefore the solution for defects on the anti-line will be shifted by $\pi$ with respect to the solution on the domain line for the same $x$.

If the line and anti-line were independent, this would cause no difficulty. The situation is different if the lines somewhat overlap. Having an overlap means that in the overlap region the system does not reach the exact vacuum state, i.e. instead of a target space pole it is in a state somewhere near that pole. This means having a non-zero $\varphi$ in the overlap region. But for a non-zero $\varphi$ we cannot have two solutions at the same time, shifted by $\pi$ with respect to each other. Therefore if $U(1)$ defects appear on a domain line, there will be no defects on the adjacent anti-line.

Now assume that we have another domain line next to the anti-line. Since there are no defects on the anti-line, there is nothing that prevents creation of defects on this second line. And since creation of $U(1)$ defects is energetically favorable, they will appear again. This is exactly the result obtained in \cite{Nagase} (see Fig. \ref{n1}): alternating lines with defects and lines with no defects.

If we have two domain lines with defects and an anti-line with no defects between them, then the anti-line makes the two series of defects independent of  each other. If the lines are parallel to the $y$ axis, this means that for a fixed $y$ they can have different $\varphi$, i.e. one can have an arbitrary phase shift with respect to the other one. This is what we see in the experimental results: series of defects on parallel lines are not aligned. Therefore we can talk about the actual crystals only with respect to a one separate line; in the other direction periodicity is not necessary and can appear only accidentally.

The absence of defects on anti-lines provides another proof of the fact that the $U(1)$ defects appear only in chains and are not isolated. Indeed, if the defects are separate then a defect on an anti-line can fit between two defects on a domain line, like depicted in Fig. \ref{Spheremapfig2} (b). Since such configurations are not observed, we can conclude that there are no significant gaps between defects.

\section{Stability}\label{StabilitySection}

The considerations in the previous sections were derived assuming no dependence of the fields on the transverse direction $z$. So it is important to understand how the finite width in the $z$ direction would affect the stability of the system. For a spherically symmetric DM interaction we can extend the known solution along the $z$ direction without $z$ dependence. But if all the configurations have the same energy, winding can happen not only in the direction of the $y$ axis. It can also happen along some direction in the $yz$ plane that is tilted with respect to the $y$ axis. Existence of equally favorable solutions along different directions leads to appearance of transverse zero modes. Interacting with the boundaries of the film, those modes may make the system unstable.

The situation is in some sense similar to the original soliton, i.e. a solitary wave in water \cite{Russell} (for a review see \cite{Bogdanov}). It can be created only in shallow water, when the close seabed prevents dissipation of the wave into the depth. In our case instead of shallow water we have a thin film, and instead of a narrow channel in which solitons are usually created we have a domain line.

Let us look at the case when the DM term is not spherically symmetric. This will happen, for instance, when
\beq
C^{k\mu} = c_k\delta^{k\mu}
\eeq
and the coefficients $c_k$ are different for different $k$. In that case if we ago along the $U(1)$ circle on the line there will be energetically unfavourable regions. Then for a system of finite width in those unfavorable regions it will become energetically favourable to change the profile to the smaller energy regions along the $z$ direction. This would provide another source of non-uniformity in the $z$ direction whose interaction with the film boundaries would endanger the stability of the system.

\section{Conclusions}

Based on the recent experimental results, we investigated the $U(1)$ defects on domain lines on thin films. We investigated their topology and showed that those defects appear due to a $U(1)$ modulus corresponding to the target space equator while the domain line connects the two vacua in the target space poles. Using the known results on the twisted mass system, we solved the model in the adiabatic approximation and got the sine-Gordon effective theory. Unlike previous theoretical considerations, we argued that it would be favorable to merge different sine-Gordon kinks into one system with a uniform winding. This issue might need further exploration.

We also considered quantization of the model, the structure of the defects on several parallel domain lines and the possible effects of non-zero width in the transverse ($z$) direction. We found that appearance of defects on a domain line prevents their creation on the adjacent anti-line. Our results on defects appearing in chains and on defect structure on parallel lines are in exact correspondence with experiment.

\section{Acknowledgements}

This work is supported in part by DOE grant DE-SC0011842.

\end{document}